\documentstyle[12pt]{article}
\newcommand{\beq}{\begin{equation}}
\newcommand{\eeq}{\end{equation}}
\newcommand{\beqa}{\begin{eqnarray}}
\newcommand{\eeqa}{\end{eqnarray}}
\newcommand{\beqar}{\begin{eqnarray*}}
\newcommand{\eeqar}{\end{eqnarray*}}

\begin{document}
\begin{titlepage}
\vspace{.5in}
\thispagestyle{empty}
\begin{center}
{\bf\Large  Aharonov-Bohm Type Forces Between Magnetic Fluxons}

\vspace{.4in}

Y. Aharonov$^{(a,b)}$\footnote{Also in a visiting position at
Boston University.},
S. Nussinov$^{(a)}$\footnote{e-mail: nussinov@ccsg.tau.ac.il},
S. Popescu$^{(a)}$,
and B. Reznik$^{(c)}$\footnote{\it e-mail: reznik@physics.ubc.ca}\\
\medskip
{(a) \small \it School of Physics and Astronomy}\\
{\small \it Tel Aviv University, Tel Aviv 69978, Israel.}\\
\medskip
{(b)\small\it Department of Physics}\\
{\small \it University of South Carolina, Columbia, SC 29208.}\\
\medskip
{(c) \small\it  Department of Physics}\\
{\small \it University of British Columbia}\\
{\small\it 6224 Agricultural Rd. Vancouver, B.C., Canada
V6T1Z1}\\
\end{center}
\vspace{.5in}
\begin{center}
\begin{minipage}{5in}
\begin{center}
{\large\bf Abstract}
\end{center}
{\small
Forces related to A-B phases between fluxons with $\Phi=\alpha\Phi_0\ \ \ $
$\alpha\ne integer$ are discussed.
We find a $\alpha^2\ln(r)$ type interaction screened on a scale $\lambda_s$.
The forces exist only when the fluxons are actually immersed in the region
with non vanishing charge density and are periodic in  $\alpha$.
We briefly comment on the problem of observing such forces.

}
\end{minipage}
\end{center}
\end{titlepage}

There is no magnetic field outside an ideal infinitely long and thin
solenoid and there is no electromagnetic force per unit length between such
parallel solenoids or fluxons.
In the following we note that the presence of charged particles between and
around the fluxons induces a new type of force between them which is
of some theoretical interest. We also briefly comment on the prospects of
detecting such interactions.

Let us assume that $n_F$ fluxons $\Phi_1,\Phi_2,....,\Phi_{n_F}$ have been
introduced at locations $\vec R_1,\vec R_2,...,\vec R_{n_F}$
where the wave function
of a system of $N$ charges $\Psi^{(0)}(\vec r_1, \vec r_2,...,\vec r_N)$ is
non vanishing.
The modification of the Schr\"odinger equation  via $\vec\partial_i\ \to \
\vec\partial_i + {e\over c}\vec A_i$ will then change the wave function
\beq
\Psi^{(0)}(\vec r_1, \vec r_2,...,\vec r_N) \ \to \
\Psi^{(0)}(\vec r_1, \vec r_2,...,\vec r_N;
\vec R_1, \Phi_1, \vec R_2,\Phi_2,...,\vec R_{n_F}, \Phi_{n_F})
\eeq
and shift the initial ground state energy $E^{(0)}$  to
\beq
E^{(0)} \ \to \ E^{(0)} +\delta E^{(0)}(\vec R_1, \Phi_1, \vec R_2,\Phi_2,
...,\vec R_{n_F}, \Phi_{n_F}).
\eeq
This induced  energy shift  can be viewed as an interaction energy between
the fluxons:
\beq
\delta E^{(0)}(\vec R_j, \Phi_j) = W(\vec R_j, \Phi_j)
\eeq
The gradients $\nabla_{\vec R_i} W$ will then yield  forces $\vec F_i$
acting on the fluxons $\Phi_i$.
To simplify the following we assume that the ground state wave function
factorizes:
\beq
\Psi^{(0)}(\vec r_1, \vec r_2, ...,\vec r_N)
=\psi_{\gamma_1}(\vec r_1)\psi_{\gamma_2}(\vec r_2)...\psi_{\gamma_N}(\vec
r_N)
\eeq
The ground state energy shift induced by introducing  the fluxons is then
a sum:
\beq
\delta E^{(0)}(\vec R_j, \Phi_j) = \sum_{i=1}^N
\delta E_{\gamma_i}(\vec R_j, \Phi_j)
\eeq
over the shifts of the individual $N$ energies $E_{\gamma_i}$ of the
$N$ charges. The latter will now be assumed to be fermions say electrons, and
suppose, for simplicity, that the electrons do not interact with each other.

Let us first consider the simplest case of  one fluxon $\Phi=\alpha\Phi_0$
introduced in the center of a cylindrical region of radius $R$.
In this geometry the free wave functions will be eigenfunctions of the
($\hat z$ component of)
angular momentum denoted by $l$ and which normally takes
on positive and negative integer values $l=0, \pm1, \pm2, \pm3,...$.
If the radial degrees of freedom were frozen then we would have simply
$E^{(0)}_{l,r} = {\hbar^2|l|^2\over 2mr^2}$. The introduction of the
fluxon effectively shifts up by $\alpha$ all the $l$ values
\beq
|l|\to |l|+\alpha \ \ for  \ \ l \ge 0, \ \ \
|l| \to |l|-\alpha \ \ for \ \ l < 0
\eeq
The {\it sum} of the energy shifts of the {\it pair}
of levels $l=\pm |l|$ is then
\beq
\delta E^{(0)}_{|l|,r} = {\hbar^2\alpha^2\over 2mr^2}
\eeq
The total energy shift is found by summing $\delta E^{(0)}_{|l|,r}$
over $l$ and $r$ values.
Bohr-Somerfeld quantization suggests discrete $r=r_n$ orbits which along
with the discrete $|l|$ in each annular region yield one
state per unit area $a_0^2$ with $a_0$ the typical distance between the
charged particles so as to fill up the Fermi "circle".
This parameter is related to the two dimensional ("one layer") density
by $n_2\sim a_0^{-2}$.
Thus\cite{c1}
\beq
W^{tot}_\alpha =\sum_{|l|,n}{\hbar^2\alpha^2\over 2mr_n^2}
\label{8}
\eeq
\beq
\to {\alpha^2\over2}{n_2\hbar^2\over2m}\int_0^R{2\pi rdr\over r^2}=
{\pi\over2}{\alpha^2n_2\hbar^2\over2m}\ln(R/a_0)
\eeq
is the total energy recquired in order to insert one fluxon at the center
of a cylinder of radius $R$. An extra $1/2$ factor is due to the fact that
there are only half as many $(l=\pm|l|)$ pairs.
The logaritmic dependence of $W^{tot}_\alpha(R)$ on $R$ is expected on
general grounds.
Assume that the fluxon is inserted at the center of a cylindrical hole
of radius $R_{in}$, inside a concentric annulus of external radius $R_{out}$.
The mininal substitution in the regular, symmetric, gauge
$\vec A =\alpha{\phi_0\over r}\hat e_\theta, \ \ \
\vec\partial\to \vec\partial +{e\over c}\vec A$, is such that the total
energy
\beq
\sum_\gamma\int |(\vec\partial +{e\over c}\vec A)\psi_\gamma|^2dx dy
\eeq
remains invariant if we scale $x\to \lambda x$ and $y\to \lambda y$,
provided we have a homogeneous uniform (two dimensional) density
\beq
n_2(x,y) =  \sum_\gamma|\psi_\gamma(x,y)|^2\simeq\ \ const.
\label{10}
\eeq
This implies that $W\simeq \log(R_{out}/R_{in})$
The argument holds also for  general  domains  of overall size
$R$ and any shape: only the coefficient in $W(R)\simeq c\ln(R)$ would
depend on the dimensionless ratio  characterizing the shape.

Equation  (\ref{8}) has a quadratic dependence $(\sim\alpha^2)$
on the energy on the flux. Due to the invariance of all topological
effects under $\Phi\to\Phi+n\Phi_0$, the energy is periodic
under $\alpha\to \alpha\pm n$.
Also the energetics should be invariant under time reversal which flips
magnetic fluxons: $\Phi_i \to -\Phi_i$. This together with periodicity
implies that for $1/2<\alpha<1$ the coefficient should become
$\sim(1-\alpha)^2$ (instead of $\sim\alpha^2$ in the $0<\alpha<1/2$ interval).

For the special case of semi-fluxons ($\alpha=1/2$) the Schr\"odinger
equation is invariant under this time reversal operation \cite{semi-fluxon}
 since the
flip ($\Phi_j\to -\Phi_j$) of a semi fluxon amounts to trivial shifts
by $\pm\Phi_0$. The wave function
$\Psi^{(0)}(\vec r_1,...,\vec r_N;  \vec R_j,\Phi_j)$
or the individual wave functions
$\psi^{(0)}_{\gamma_i}(\vec r_i; \vec R_j,\Phi_j)$
can be made real by  choosing the singular gauge
$A_\theta=\delta(y),\ \ x>0$ for a
fluxon at the origin.
In this case real wave function $\psi(x,y)$ simply jumps along the $x>0$ ray,
$\psi(x,y=+\epsilon) = - \psi(x,y=-\epsilon); \ \ \ x>0$,
consistent with the requisite A-B phase of $\pi$ picked upon
circulating the fluxon.
Thus real wave function flips sign an odd number of times along
any closed path which encloses only $\Phi_1$ at $\vec R_1$.
Therefore an odd
number of continuous strings of zeros emanate from
$\vec R_1$, and from all other
fluxons at $\vec R_j$. These  "null lines"
\cite{nl} \cite{semi-fluxon}
 terminate on the boundary of the system
(beyond which $\psi_{\alpha_i}(x,y)$ vanish anyway) or,
at the location of  another fluxon.

In the case of cylindrical symmetry the intial angular real wave functions
($\sin(|l|\theta), \ \ \cos(|l|\theta)$) have each $2|l|$ nodal
lines at the origin.
The shifts $|l|\to |l|\pm {1\over2}$, induced by introducing a semi-fluxon
at the origin, are equivalent to adding/substructing one null line -
yielding  in both cases an odd -  $(2l\pm1)$, number of null lines as
required by the general considerations.

Returning to the main issue let us address next the mutual interaction
energies of two semi-fluxons at a distance $|\vec R_1-\vec R_2|=a$
inside a uniform medium of charged particles, with constant two dimensional
density $n_2$ give by equation (\ref{10}), introduced near the
origin at the center of a large domain.
At a distance $r$ from the origin the vector potentials are:
 $\vec A_i = {1\over2}
\Phi_0 {\hat e_{\theta_i}\over |\vec r- \vec R_i|}\ \ i=1,2$
where $\hat e_{\theta_i}$ refer to the tangential unit vector with
respect to $\vec R_1$, $\vec R_2$ as origins. For $r>a$ $\vec A_1+\vec A_2$
add, up to small corrections, to the vector potential of a single quantized
fluxon at the origin. Insofar as the topological effect of
interest are concerned the latter is
just a gauge artifact. Hence we expect that
only a region of size of order $a$ around each fluxon and between
the fluxons will be affected and consequently that mutual interaction energy
behave like $W^{tot}(r)$ of eq. (\ref{8}) with $r\sim a$.:
\beq
W_{(\alpha_1=1/2,\alpha_2=1/2)} (a) = \xi {\pi\over16}{n_2\hbar^2\over m}
\ln(a/a_0)
\label{12}
\eeq
The numerical factor of order one $\xi$ representns the effects of having
a two center system.

This interaction leads then to an attractive force between the two
semi-fluxons
\beq
F_{(1/2,1/2)}(a) \simeq {\xi\pi\over 16}{n_2\hbar^2\over m} {1\over a}
\eeq
Recalling that $W$ and $F$ represent the effect of "one layer"
and $n_2$ is the two dimensional number density in this layer, we
can rewrite the last equation in a more usfull form as
\beq
{F_{(1/2,1/2)}(a)\over unit\ \ fluxon\ \ length} \simeq
{\xi\pi\over 16} {n\hbar^2\over m} {1\over a}
\label{15}
\eeq
with $n$ the true three dimensional density of the charges.
The coefficient $\xi(\alpha_1,\alpha_2)$ should be periodic
in both $\alpha_1$ and $\alpha_2$. For positive $\alpha_i$
the force will be attractive if
$\alpha_1+\alpha_2>1/2$. However for $\alpha_1+\alpha_2 <1/2$
we expect repulsion: the energy of the
joint system with $\Phi_1 \ \ \Phi_2$ overlapping ($\sim
(\alpha_1+\alpha_2)^2$)
exceeds the sum of energies of two seperate fluxons: $\sim
 \alpha_1^2+\alpha_2^2 $.

The topological origin of the forces is clearly manifest by the fact that
such forces act only on fluxons which are actually immersed in the charged
particle background but is absent for fluxons which are outside this region.
Thus if in the example of the concentric cylindrical geometry discussed above,
we move the fluxon inside a hole, the energy of the system is unchanged
and no force is expected. Since there are no charged particles in
the hole we can continue using the same gauge potential $\vec A(r)
={\alpha\Phi_0\over r}\hat e_{\theta}$ even when the fluxon is not
in the center. The key point is  for every path enclosing the fluxon (
or fluxons -
if there are several fluxons inside the hole) that a charge particle confined
to the anular domain can {\it actually} perform - the AB phase will be the
same.

The interaction (\ref{12}) is quasi - confining ($W(r)\to \infty $ with
$r\to \infty$) just like the two dimensional coulomb interaction.
It is well known that for such cases the system may find it energetically
favorable once $R\ge \lambda_{s}= (screening \ \ length)$, to screen the
charges (or fluxons in the present case). Indeed such a screening
 is generated by the circulation of all the charged particles of charge $e$
(for fluxon of $\Phi=\Phi_0/2$ say).
The corresponding current density at a distance $r$
is
\beq
\vec J(r) = {\hbar \alpha n\over m r } \hat e_{\theta}
\eeq
The screening of the $B$ field is found from Maxwell's equation:
\beq
{dB_z^{induced}\over dr} = {1\over c}J_\theta = {\alpha e \hbar n\over
mcr}
\label{16}
\eeq
The $\alpha$ in eq. (\ref{16}) depends on $r$ due
to the partial screening of the initial fluxon $\alpha=\alpha(r=0)$
by currents circulating between the origin and $r$:
\beq
\alpha(r) = (\alpha\Phi_0 - \int_0^r2\pi rB^{induced}_z(r)dr)/\Phi_0
\label{17}
\eeq
The coupled equations (\ref{16}) (\ref{17}) yield  $r$ profiles
for $\alpha(r)$ and $B_z^{induced}(r)$  which are exponentially falling off
 like $\exp(-r/\lambda_s)$, thus defining $\lambda_s$.
Approximating $\alpha(r) =\alpha\theta(\lambda_s-r)$
we readily find $\lambda_s$ from
\beq
2\pi\int_0^{\lambda_s+\epsilon} rB_z^{induced}(r) =
\pi\int_0^{\lambda_s+\epsilon} r^2{dB_z\over dr} =
\eeq
\beq
= {\pi\over mc} {e\alpha\hbar \lambda_s^2\over 2} = \alpha\Phi_0 = {2\pi\alpha
\hbar c\over e}
\eeq
where we used integration by parts, eq. (\ref{16}), and demanded that
the net induced flux exactly cancel $\alpha\Phi_0$.
Recalling that $\alpha_{em} = {e^2\over \hbar c} = {\lambda_{comp}\over
a_{Bohr}}$
we can write $\lambda_s$ as:
\beq
\lambda_s = {2\over \alpha_{em}}a_{Bohr}\Bigl({a_{Bohr}^{-3}\over n}\Bigr)
^{1/2}=
150 \Bigl({10^{25}\over n}\Bigr)^{1/2} A^0
\label{21}
\eeq

We considered so far a system of free charged fermions. The induced
interactions are actually the same for charged
bosons. In this case
we need not fill up a Fermi sphere of levels and at $T=0$ all bosons would
be in the same ground state $\psi_0(x,y)$.
However from the above derivation each occupied level $\gamma_i$
contributes equally and the same $W(R)$  emerges when
all bosons are in the ground state.
This is true also for a  superconductor
in the Landau-Ginzburg effective charged Higgs model description.
We then find an extra
energy $\simeq \int d^2r |(\vec\partial - {e\over c}\vec A)\phi(r)|^2
\simeq \alpha^2\phi_0^2\ln(R/R_0)$ where $\phi_0$ is the
order parameter due to an improperly quantized fluxon
leading again to a $1/r$ force.

In passing we note that
QCD is a nice example of confinement-screening interplay.
The spectra of heavy quark-antiquark, $\bar Q Q$, system suggests a
confining linear $\bar QQ$ potential $V=\sigma R$ at "large" distances and
the same is expected to $QQ$ baryons in $SU(2)_c$. Creation of $\bar qq$
pairs tends to screen the confining potential - and only exponenetially
falling Yukawa like potentials exist between physical, color neutral, hadrons.

The QCD quarks with non zero  triality (screening the confining
 interaction between $\bar QQ$) play the role of the electrons
which transform non trivially under the "$Z_2$" of the fluxon in our example,
and generate currents screening the interaction  between the
semifluxons.
The mechanisems for screening and confinment tend to be mutually exlusive:
 both in QCD and in our example the screening of charges
reduce the long range forces and resulting putative pairing of $Q\bar Q$,
(two semi-fluxons here).
Also  $\bar qq$'s which are already paired by confinment to triality
(and color) singlets will not screen the $Q\bar Q $ force.
The introduction of the two semi-fluxons will not induce here large scale
pairing of the electrons. Yet the mechanism of semi-fluxons
confinement may quench the screening currents.
This could be the case for the Bose-Einstein condensate example.
If the distance between the two fluxons $a$ is smaller then the size of the
system $R$, a null line will form between them\cite{c2}.
This in turn impedes the circulation of screening currents around $\Phi_1$
or $\Phi_2$ separately.

For conducting mesoscopic rings of sizes of order of microns,
 voltages and persistent
currents related to $W(r)$ and $J$ above have  already been
observed \cite{webb}.
Can one detect also the force between fluxons?
For this the coherence length for electrons and the screening length
$\lambda_s$  of eq. (\ref{21}) should both exceed the fluxons seperation $a$.
Fot typical metals $n\simeq 3\cdot10^{22}$ and $\lambda_s\simeq 0.2\mu$.
Together with the intrinsic requirement following from the general
topological argument above that the fluxons be immersed
inside the metal this appears to make the  measuremt  of
the force of
eq. (\ref{15}) in metals virtually impossible.

Type $II$ superconductors with semi-fluxons (integer fluxons in the Copper
pair  $2e$ charge unit),  generated by super currents on
a penetration length of scale $\lambda_{p}\simeq 0.2\mu$,
appear more promising.
By varying  the temprature in the interval $T_c>T>0$
the density of normal, unpaired, electrons can be controlled.
The pattern of normal electron circulation would, on its own, be
energetically disfavored because it would generate "forbidden"
$B$ fields inside the
superconductor for $\lambda_{p}<r<\lambda_s$ and compensating
super-currents should locally cancel it.
If this does not completely "freeze the system"
the introduction of the semi-fluxons will raise its energy by amounts of
order ${W(r)\over unit\ \ length}\simeq n_e{\hbar^2\over2m_e}\ln(r/r_0)$.
 For $n_e=10^{19}$,
a $1/r$ force of order
$10^{-5}-10^{-4} \ dyne/cm$ would then operate
betwee fluxons which are $1-10\mu$ appart.
 Such forces may tend to pair semi-fluxons
or bend them. It is not inconcieveable that such effects may be detected.

Finally it is amusing to compare the topological
force (\ref{15}) with the Casimir
force \cite{Duru} between two parallel conducting wires $\sim {\hbar c\over
a^3}$.
The ratio is:
$\rho\equiv F_{top}/F_{cas} \sim n \hbar a^2/mc^2 \sim n \lambda_{com} a^2$.
For electron systems $n\sim a_0^{-3}$ with $a_0$ of the order
of the Bohr radius. Using $\lambda_{com}/a_0 \sim \alpha_{em}$ we
have then $\rho \sim {\alpha_{em}a^2/ a_0^2}$
Since generally $a>>a_0$ this ratio is very large.
The origin of this large ratio is easy to assess. Only vacuum fluctuations
(photons) on scales $\lambda\sim a$ contribute to the Casimir force
whereas {\it all} electron modes down to wavelength $\lambda \sim a_0$
contribute equally to the interaction energy and force proposed here.

\vspace{1in}

{\bf Acknowledgement}

We have greatly benefited from the help and advice of C. K. Au,
R. Creswick, H. Farach
and C. Poole and particularly from the critical comments of A. Casher.


\begin{thebibliography}{99}
\bibitem{c1}
The problem of finding $W^{tot}_\alpha(R)$ for the case when we use
the  eignfunctions $J_{l}(k^{(0)}_{n,l}r)$ with  energies
$(k^{(0)}_{n,l})^2 $ is somewhat involved. The $k_{n,l}$ are determined
implicitely from the n'th solution of $J_l(kR)=0$ which is rather
complicated for the large $l$ and $kr$
required here.
\bibitem{c2}
We found that the introduction of such a null line into states with
electrons wavelength $\lambda>a$ and a-fortiori into the ground state
with $\lambda\simeq L$ will cost only a logaritmic, (rather than linear)
energy consistent with the other derivation.
\end{thebibliography}
\end{document}